\documentstyle{article}

\newcommand{\nl}{\newline}

\title{ Are there Floquet Quanta?}

\author{Francisco Delgado C.
\thanks{Depto. de Matem\'aticas, ITESM, Campus Estado de M\'exico. \nl
A.P. 2, 52926, Atizap\'an, Edo. de Mex., M\'exico.} \\ \\
and \\ \\
Bogdan Mielnik
\thanks{Institute of Theoretical Physics, Warsaw University. \nl
 ul. Hoza 69, 00-681, Warsaw, Poland.} \\ \\
Depto. de F\'\i sica, CINVESTAV, \\
A.P. 14-740, 07000, M\'exico D.F., M\'exico.}

\date {}

\begin{document}
\begin{titlepage}

\maketitle

\begin{abstract}

The Zeldovich hypothesis is revised and  the meaning of quasi energy spectra is discussed.  The
observation of Floquet resonance for  microobjects in quickly oscillating external fields might bring
a  new  information  about  the  time   scale   of hypothetical quantum jumps.

\vskip 1cm

PACS number(s): 42.50.Lc, 33.55.Be, 33.80.Ps, 42.62.-b.  

\vskip 1cm

{\it Preprint} CINVESTAV: CIEA-FIS/97-06

\vskip 1cm

\end{abstract}

\end{titlepage}

When considering the absorption (emission) spectra one usually  has in mind a static (stationary) system. 
"In itself" (i.e.,  when  isolated from the rest of the universe) it is described  by  a  time  independent
Hamiltonian. When submerged in external fields,  however,  it  starts  to radiate:  the  differences  between  the  eigenvalues  of  the  Hamilton operator define the energies of emitted (absorbed) quanta.

The physical reality though, is not limited to  static (stationary) systems. In fact, if any physical theory was  at  all formulated, this is only since we live in a variable universe, where the
external fields can be changed and the experiments can be performed.  An intriguing question thus arises: can a  system  with  a  time  dependent Hamiltonian have a similar resonance capacities as the static systems?

A known attempt to give an answer belongs to  Zeldovich  ~\cite{Zel}  and concerns the systems with periodic Hamiltonians. The  unitary  evolution operators $U(t,t_0)$ obey:

 \begin{equation}
{dU(t,t_0) \over dt} = - i H(t) U(t,t_0)
\end{equation}                                

\begin{equation}
H(t) = H(t+T)   \quad \quad  (t \in {\bf R})
\end{equation}                                 

According to the idea of Zeldovich the properties of  the  periodic system (1-2) are determined by its Floquet operator, i.e.,  the  unitary operator $U(T)=U(T,0)$ describing the evolution within the complete period $T$. Obviously:

 \begin{equation}
U(T) = e^{-iTF}
\end{equation}                            
where the self-adjoint operator $F$ is called the Floquet Hamiltonian. The hypothesis of Zeldovich tells that the  eigenvalues  of $F$, though not energies themselves (the proposed  term  is  {\it quasienergies} ~\cite{Zel}), determine the resonance spectrum  of  the  periodic  system  (1) modulo multiples of $\hbar \omega$ (where $\omega=2 \pi / T$).

The idea, though intuitive, leaves some  questions  open.  In  the first place, the definition of $F$ is non-unique. Every $U(T)$ in (3) admits an infinity of Floquet Hamiltonians (corresponding to the $n \hbar \omega$  tolerance
in the spectrum) and it is not obvious which $F$, if any, has  the  energy interpretation. In fact, in some recently  studied  cases,  the  Floquet generator which most naturally describes the evolution, precisely cannot enter into a conservative balance with the external radiation. The first such cases were found in ~\cite{BD1,Dav}  by  observing  that  for  the  charged Schr\"odinger's particle in  a  magnetic  field  ${\bf B}(t)$ uniformly  rotating around a fixed vector ${\bf n}$  ($|{\bf n}|=1$), the evolution operator becomes:

\begin{equation}
U(t,0) = e^{-i \omega t {\bf n} \cdot {\bf M}} e^{-itF}
\end{equation}                                  
where $F$ is a linear combination of three 1-dimensional  oscillators, $F=H_1+H_2-H_3$ (one sign negative!). Every $t=nT=2n \pi / \omega$ the  first  factor  in (4) reduces to 1, and so, $F$ is a natural Floquet Hamiltonian. However, $F$ cannot be the right counterpart for the radiative energy (otherwise, the system could emit an infinite energy at the cost of  falling  down  into negative $F$-levels).  A  similar  phenomenon  occurs  for  the  molecular rotator whose electron states resemble the 'Troyan asteroids' ~\cite{IBB1}. In
both  cases  the  'quasi  energy   crash'   excludes   a   good   energy interpretation for $F$. Note, that  the  difficulty,  apparently,  escaped the attention of Zeldovich himself, who wrote about the "..transitions  from
the lowest quasienergy eigenstate (...) into an excited  state..."  (see ~\cite{Zel}, p.1007). A part of the problem is attended in the new study of  the "Troyan case" ~\cite{IBB2} (the resonance hypothesis  of  Zeldovich  ~\cite{Zel}  is confirmed by the first order perturbation, though the stability  of  the ground-top state is still an open problem).

The phenomenon of the 'top state' is not the only puzzle. In  fact, in some simple models the insufficiency of the  Floquet  Hamiltonian  to describe the {\it complete resonance} is  immediately  obvious.  The  simplest
case occurs if $H(t)$ is a periodic operator-valued step-function taking a finite number of steps:

\begin{equation}
H(t)=H_1, H_2, ... H_n \quad \quad (periodic \quad pattern)
\end{equation}                        
in time lapses $\tau_1, \tau_2, ... \tau_n$ ($T=\tau_1+...+\tau_n$  being the  $H(t)$-period). The  Floquet  Hamiltonian  $F$  then  is  the  Baker-Campbell-Hausdorff exponent:

\begin{equation}
e^{-i \tau _n H_n}...e^{-i \tau _1 H_1}=e^{-i (\tau _1 + ...+ \tau_n) F}
\end{equation}                               

According to  the  quasi-energy  hypothesis  ~\cite{Zel},  $F$  should define the radiation spectrum for the periodic process (5).  This  seems true if the jumps in (5) are very fast ($T$ small). However, if  the  time
lapses $\tau_1,...,\tau_n$ are long enough comparing with the  typical  absorption (emission) time, then the absorbed (emitted) quanta  will  'see'  either only $H_1$, or $H_2$, etc, without 'noticing'  $F$.  In  this  way,  the  Floquet spectrum is linked with a deeper question about the  effective  time  of the absorption (emission) processes. The problem is intimately related to the {\it epicycle structure} of the evolution operator $U(t,0)$. Put $G(t,0)=U(t,0) e^{itF} \quad \Rightarrow$

\begin{equation}
U(t,0)=G(t,0) e^{-itF}
\end{equation}  

where $e^{-itF}$ represents the "main evolution trend" while $G(t,0)$ is a 'closed loop operation' returning to 1 for every $t=nT$. The operator (7), in general, does not allow for the stationary states, though it permits the existence of periodic ones. Indeed, suppose $F$ has a point-spectrum with a sequence of eigenvectors $\phi_1, \phi_2, ...$ belonging to the eigenvalues $\omega_1, \omega_2,...$ The state trajectories $\phi_n(t)$ originated by $\phi_n$'s then are:

\begin{equation}
\phi_n(t)=U(t,0) \phi_n = G(t,0) e^{-itF} \phi_n = G(t,0) e^{-i \omega_n t} \phi_n
\end{equation} 

The assumptions of Zeldovich ~\cite{Zel} mean that the Floquet photon does not interact at all with the "circulating part" $G(t,0)$; it penetrates "right to the bottom" of the dynamical process (8), where it simply replaces $\phi_n$ by $\phi_m$; the loop operator $G(t,0)$ acts as before. 
To have an  exactly  soluble  model, consider the 1-dimensional oscillator:

\begin{equation}
H(t) = {p^2 \over 2} + \beta(t)^2 {q^2 \over 2}
\end{equation}                                 
where $\beta(t)$ is a periodic function. The epicycle structure is most regular if the Hamiltonian (9) causes an  'evolution loop' (a process in which all motion trajectories simultaneously close,
and the entire $U(\tau,0)$ turns  proportional  to  1  after  a  finite number of  periods  $\tau=nT $ ~\cite{Loop1}).  The  quasi-energy spacing of the loop is $ \Delta F=\hbar \omega_F$, where  $\omega_F=  2 \pi l/\tau$ ($l=0,\pm1,\pm2,...$). The general cases of Floquet  spectra  has  been  most
carefully   studied   for    the    rotating    fields    ~\cite{Paul, BD1,Dav,IBB1,IBB2};  the  oscillating  case  elaborated  for  ion  traps whenever the use of Mathieu functions was accessible ~\cite{Paul};  the  exact
numerical study of more general cases is still fragmentary. Our Fig.1 plots the numerically  determined  Floquet  frequencies  $\omega_F$  for  rectangular  and sinusoidal $\beta(t)$. The  loop  processes  occur  whenever  $\omega_F$  crosses  the
multiple of $2 \pi / nT$; in all cases the Floquet photon "feels" only the global form of the trajectory (7-8).

Note that the pulsating systems (9) can be produced in laboratory if $\beta(t)$ corresponds to the  intensity  of  a  homogeneous,  time  dependent magnetic field ~\cite{Loop1,BD1} of  a  cyllindrical  solenoid,
${\bf B}(t)={\bf n} B(t)$, $B(t+T)=B(t)$ ({\bf n} is a unit vector).  The  vector  potential  is  ${\bf A}({\bf x},t)=(1/2) {\bf x} \times {\bf B}  = (1/2) B(t) {\bf x} \times {\bf n}$ and the Schrodinger's particle of charge $e$ and mass $m$ obeys the Hamiltonian:
                         
\begin{equation}
H(t)={1 \over 2m} [{\bf p}-{e \over c} {\bf A}]^2
\end{equation}    

or in the simplified variables  $q_1=x \sqrt{m}/ \hbar$,  $q_2=y \sqrt{m}/ \hbar$,  $q_3=z \sqrt{m}/ \hbar$, $p_1=p_x/\sqrt{m}$, $p_2=p_y/\sqrt{m}$, $p_3=p_z/\sqrt{m}$ :

\begin{equation}
H(t)={1 \over 2} p_3^2 - \beta (t) M_3 + [{p_1^2 \over 2}+{p_2^2 \over 2}+ \beta (t)^2 ({q_1^2 \over 2}+{q_2^2 \over 2})]
\end{equation}                              
where the axes $x$, $y$, $z$ are respectively orthogonal or parallel to the unit vector ${\bf n}$ and the 'manipulation function' $\beta(t)=e\hbar B(t)/2mc$  can  simulate pulses of any shape in (9). The Floquet phenomenon (10-11) too, has its extremely regular forms. Thus e.g., Fig.2 represents a loop case generated by 24 periods of the sinusoidal field :

\begin{equation}
{\bf B}(t) = {\bf n} B_0 \sin \omega t, \quad \quad |{\bf n}|=1
\end{equation}                                  
How does such a system interact  with  an  external radiation? While the resonant response to coherent fields of Floquet frequency is beyond any doubt (compare the semiclassical aproach of Rabi et al. ~\cite{Rabi}) the research on magnetic resonance might indicate the domination of  multi-photon processes ~\cite{Mult1}. The  absorption (emission) of single quanta is a distinct phenomenon leading to  some  less typical problems.

The {\it time of events} in quantum theory is  the  subject  of  unfinished  discussions  ~\cite{Tate}. The question about the {\it minimal time} for an act of absorption
(emission) is seldom adressed (if not discouraged) by  the  present  day formalism. By applying the quantum equations 'to the  letter', one might conclude that the  emission,  absorption,  decay  are  virtual
processes, never indeed concluded. Opposite   arguments
(returning to the pionieer ideas!) indicate, that the acts of absorption (emission) {\it indeed happen} ~\cite{Cook,Jum1}: they  are  sudden  jumps from 'potentiality' to 'actuality'; a kind of spontaneous reductions  of quantum state, leaving  no  slightest  doubt  that  the  absorption  {\it has 
occured!} In the lab scale the jumps, apparently, are not  restricted  by any  {\it minimal time}  (though  the  {\it expected  time}  might be finite:   see the anti-bounching  phenomenon  ~\cite{Anti1}).  Would  the  picture  be similar for the Floquet absorption?

To avoid 'doctrinal constrains' we shall stick to intuitive ideas ~\cite{Cook}. Assume that a photon penetrates into  a  solenoid where a microobject is kept under the influence  of  a  magnetic  2-step pattern: $B_1, B_2, B_1, B_2,...$ Let the magnetic steps $B_1, B_2$ last 1 min. each, with  $T=2$ min.  Of course, within the first 1 min. the  system  can  absorb  any  photon  of energy $\Delta E_1= \hbar \omega_1$ whereas during the next 1 min. it can absorb  any  photon  of energy  $\Delta E_2=\hbar \omega_2 $  ($\omega_i=e B_i/2mc, \quad i=1,2$).  In contrast, the absorption of a 'Floquet photon' is  the summary effect of the entire period of  $H(t)$: so,  it  should  not  occur until the magnetic field indeed accomplish  the  2-pulse  pattern.  (The best argument is the {\it reductio ad absurdum}. If the  mechanism  generating the double $B(t)$ pulse had a sudden defect and if $B(t)$ failed to  produce the 2-nd step $B_2$, the Floquet frequency would  never  be  absorbed! Should the Floquet photon be absorbed during the first step,  how  could it 'know' that the second step will indeed occur?) This suggests the {\it minimal time} $T=2 min.$ needed for the 'Floquet absorption'. Quite similarly, for the general process (7-8) the photon would have to wait until the "loop evolution"  $G(t,0)$ closes up, to 'see' the global aspect $e^{-itF}$ behind. Can the single photon absorption be so incredibly slow?

To find an answer, the only method is an experiment: one has to place a sample of  identical  quantum objects in the oscillating magnetic field (10-12). The sample should  be then bombarded by an external  photon  beam  of  Floquet  frequency.  To
distinguish the Floquet resonance to single quanta from  the  parametric resonance to coherent fields (which can  involve  multiphoton  processes ~\cite{Mult1}) it might be necessary to apply monocromatic but perfectly incoherent photon beams, so that the photons drop  separately  onto  the sample. (To create such a beam  is  a  separate  challenge  but  is  not fundamentally impossible). The resonance absorption should be also checked for the "instantaneous spectra" of $H(t)$. Now, the exclusive presence of the (diffused) instantaneous levels of $H(t)$ would mean that the acts of absorption are much quicker than the  period of the external field (the absorbed photons have simply no time  to  get involved in the Floquet process). In turn, the appearence of the sharply defined 'Floquet lines' would confirm the existence of 'slow absorption' correcting the ideas about quantum jumps ~\cite{Cook}. An analogous conclusion should hold for the 'Troyan rotators' ~\cite{IBB1, IBB2}.

An essential difficulty are  orders  of  magnitude. The  pulsating fields of electromagnets might turn 'too slow' to make the problem  more than academic (for low $\omega_F$ it might be impossible to bombard  the  sample with single Floquet quanta!) If faster, the  system  would  generate  an extra radiation spoiling the approximation (9). Note though,  that  clean and fast oscillating fields operate in the nodal  points  of  the  laser beam traps ~\cite{Laz1,Laz2,Laz3}.  For  powerful laser beams $\simeq 10^{15} Watt/cm^2$ the magnetic fields in principle, can approach $10^6 G$, comparable  to  the  newest  achievements  of  the macroscopic  technology  ~\cite{Magn}. (Indeed, we find it strange that the laser beam  traps  are  so seldom used; they  might  mark  some  natural  time  scales  for  the  atomic phenomena!).  Thus,  e.g.,  two  monocromatic,  perpendicularly  crossed standing waves, have the vector potential:

\begin{equation}
{\bf A}_{\bf mn} ({\bf x},t) = {1 \over 2} A [{\bf m} \sin({\omega \over c} {\bf n} \cdot {\bf x}) - {\bf n} \sin({\omega \over c} {\bf m} \cdot {\bf x}) ] \sin(\omega t) 
\end{equation}              
({\bf n},{\bf m} ,{\bf s} are three orthogonal unit vectors) hosting the  sinusoidally  pulsating  field  (12) on the nodal  line ${\bf m} \cdot {\bf x}={\bf n} \cdot {\bf x}=0$  ~\cite{BD2}.  The equivalence to the 'solenoid model' (10-11) and to the oscillator (9) is
local; but it should hold as long as the charged particle  is  mantained in vicinity of the  nodal  line  (the  typical wavelenghts of the lasers are $\simeq 10^{-6} m$, while the atomic size $\simeq 10^{-9}  m$). The stability thresholds are  another obstacle  (though  only  for  charged microobjects). To keep a  charged  particle  in  the  oscillating  field (12-13)  the  ratio  of  the  amplitude/frequency  cannot  be  too  high
(otherwise  the  particle is expulsed  ~\cite{BD1,BD2}).   For   the sinusoidal pulses (12), the crucial parameter  is $\alpha=eB/2mc \omega$ and the stability condition is ~\cite{BD2}:

\begin{equation}
|\alpha| < 0.5735...
\end{equation}                                           

For a neutron, there is no threshold (14),  and  the  new techniques ~\cite{Magn} permit to apply  strong  fields  to  examine  the  Floquet
spectra. In case of the oscillating laser fields the situation  is even better due to the high frequencies. Of course, to create a high intensity standing wave with an  exact  nodal line  (13)  is  a  non-trivial  task  (but  must  all  efforts  of   the experimental physics be always dedicated to particle accelerating?)

An interesting class  of  traps  is  obtained  by  superposing  two standing waves $A_{\bf ms}$, $A_{\bf ns}$ so that the nodal  lines  intersect  and the phase difference is $\pi/2$. The resulting field  has  a  net  of  nodal
points hosting the rotating magnetic field ~\cite{BD1}:

\begin{eqnarray}
{\bf A}_{\rm rot} ({\bf x},t) & = & {A \over 2} [{\bf m} \sin({\omega \over c} {\bf n} \cdot {\bf x}) - {\bf n} \sin({\omega \over c} {\bf m} \cdot {\bf x}) ] \cos(\omega t) \nonumber \\
  &   & + {A \over 2} [{\bf s} \sin({\omega \over c} {\bf n} \cdot {\bf x}) - {\bf n} \sin({\omega \over c} {\bf s} \cdot {\bf x}) ] \sin(\omega t) \\
 & \matrix{\simeq \cr _{{\bf x} \rightarrow 0} \cr} & {1 \over 2} {\bf x} \times {\bf B} (t); \quad \quad {\bf B} (t) = {A \omega \over c} [ {\bf m} \cos \omega t + {\bf s} \sin \omega t] \nonumber  
\end{eqnarray}                  

To assure that the nodal lines of two standing waves (13) intersect exactly is again a formidable challenge  -  but  if  achieved  it  would permit to observe the effects  of  strong  and  fast  rotating  magnetic fields in micro scale. An interesting experiment would  be  to place a spin $1/2$ particle (electron, neutron) in the  rotating  magnetic field (15) and check for the magnetic resonance (not to $\omega$  as  described in  ~\cite{Rabi,Mult1} but  to  the   Floquet   frequency   $\omega_F$ !).   The 'instantaneous Hamiltonian' is:

\begin{equation}
H(t) = - \mu  {\bf B} (t) \cdot {\bf \sigma}
\end{equation}                                           

   The transition to the 'rotating  frame'  ~\cite{Rabi,BD1,IBB2} yields  the evolution operator:

\begin{equation}
U(t,0)=e^{-i \omega t {\bf n} \cdot {\bf \sigma} /2} e^{-itF}
\end{equation}                                      
where $F$ is the new time independent Hamiltonian and  simultaneously, the most natural Floquet generator for (16):

\begin{equation}
F = -\mu \hbar B \sigma_x + \omega \hbar \sigma_z /2 = \left( \matrix{ \hbar \omega/2 & -\mu B \cr -\mu B & - \hbar \omega /2 \cr} \right) 
\end{equation}                                        
with two eigenvalues: $\lambda_{\pm}  =  \pm  \sqrt{  \mu^2 B^2 +  (\hbar \omega /2)^2}$  [independent  of  the particular representation (18)]. As already noticed, $F$ may have no  good energy interpretation (compare  ~\cite{BD1}).  This  indeed  happens  for  the generator (18) which conserves a non-trivial spectrum for $B \rightarrow 0$. Knowing that the quasi-energies are defined modulo $\hbar \omega$  one  immediately  gets  the   right   spacing   for   the   magnetic Floquet-resonance of (16):

\begin{equation}
\Delta E = \hbar \omega ( \sqrt{1 + ({2 \mu B \over \hbar \omega})^2}-1)
\end{equation}                          
with the correct limiting values:

\begin{eqnarray}
\Delta E & \simeq & 2 \mu B {\mu B \over \hbar \omega}, \quad \quad   2 \mu B << \hbar \omega \nonumber \\
         & \simeq & 2 \mu B - \hbar \omega, \quad \quad  
\hbar \omega >> 2 \mu B 
\end{eqnarray}                               

As before, the observation of the magnetic Floquet line (19-20) would mean the existence of the slow absorption with $\tau  >  2 \pi/\omega$,  whereas  the
domination of the instantaneous line with $\Delta E=2 \mu B$ would testify that the absorption (emission) times are much shorter than the  trap  oscillation period $T=\omega /2 \pi$.

An important experiment would be to check the Zeeman and magnetic resonances  for microsystems  in  the  linearly  oscillating  fields.  Suppose, a microobject  with a spherically symmetric Hamiltonian $H_0$ is kept in the field (10-12). If the terms quadratic in $B(t)$ are negligible (approximately true  for  Zeemann
if  $B \simeq 10^4 G$; exactly for the magnetic resonance ~\cite{Mult1}), the  Floquet  Hamiltonian  $F$  is  identical  with  the unperturbed  $H_0$  (the  contributions  from  $-{\bf M} \cdot {\bf B}(t)$  cancel. The conclusion holds also for the Anandan-Hagen term ~\cite{Aha}). In contrast, the 
instantaneous Hamiltonians $H(t) = H_0 - {\bf M} \cdot {\bf B}(t)$ should show the  (variable) Zeemann  spectra.  Assume  now,  a  sample  of microobjects in  the  field  (12)  is  additionally  bombarded  by  an incoherent photon beam. Then, the existence of ordinary spectral  terms, without Zeemann corrections would mean the  domination  of  the  Floquet mechanism (slow absorption). Should the Floquet  spectrum desappear for too low $\omega$ (slow field oscillations),  it  would  mean  the
absence of too slow absorption- emission acts.  To  the  contrary,  the "diffuse lines" corrected by $ - {\bf M} \cdot {\bf B}(t)$ (for  variable  $B(t)$)  will  mean  'quick' emission- absorption processes, confined to very  short  time  intervals
(the conclusion seems valid even if the 'separately dropping photons are not available!) Henceforth, the absence of such lines for high $\omega$,  could mean the existence of a {\it minimal time} for absorption  blinding  the vision of the 'instantaneous Hamiltonians' $H(t)$. If  there  is a   minimal absorption time of few nanoseconds compatible with the antibounching observations ~\cite{Anti1}, then even lower frequencies can be used to blind the instantaneous spectra of $H(t)$.  

{\bf Acknowledgements.} The support of CONACYT, M\'exico, is acknowledged.

\newpage

\centerline{\bf Figure Captions}
\vskip 1cm

{\bf FIG. 1 .-} The spectral Floquet frequency $\omega_F$ for an oscillator (7) driven by: ({\it a}) time independent $\beta (t) = \beta_0$; ({\it b}) by a sequence of rectangular pulses $\beta(t)=\beta_0, 0, \beta_0, 0,...$ in the time lapses $T/2, T/2, T/2,...$; ({\it c}) by the sinusoidally varying $\beta(t)=\beta_0 \sin \omega t$. Below, three cases of uncharacteristic epicycles (in form of closed loops) generated on the phase plane: ({\it a}) the constant $\beta_0=\beta_{0a}=1.57079...$; ({\it b}) by rectangular pulses with $\beta_0=\beta_{0b}=2.15375...$; ({\it c}) by the sinusoidal pulses with $\beta_0=\beta_{0c}=2.21231...$

\vskip 1cm

{\bf FIG. 2 .-} Specially regular Floquet process in form of an evolution loop generated by 24 periods of the sinusoidal magnetic field ${\bf B}(t)=(B_0+B_1 \sin \omega t) {\bf n}$ with $\beta_0=0.78539..., \quad \beta_1=0.94595...$ in the plane orthogonal to {\bf n}. The loop effect is shared by the classical  and  quantum motions. How does the phenomenon interact with an external radiation?

\end{document}